\documentclass{emulateapj}
\usepackage[dvips]{color}
\usepackage{amssymb, amsmath, amsbsy, epsfig, epsf, slashbox}

\newcommand{\lbol}{\ensuremath{L\mathrm{_{bol}}}}

\newcommand{\ledd}{\ensuremath{L\mathrm{_{Edd}}}}
\newcommand{\lratio}{\ensuremath{L/\ledd}}

\newcommand{\lfive}{\ensuremath{\lambda L_{\lambda}(5100)}}

\newcommand{\msun}{\ensuremath{M_{\odot}}}

\newcommand{\mbh}{\ensuremath{M_\mathrm{BH}}}

\newcommand{\pnull}{\ensuremath{P_{\mathrm{null}}}}

\newcommand{\ha}{{\rm H\ensuremath{\alpha}}}
\newcommand{\hb}{{\rm H\ensuremath{\beta}}}

\newcommand{\oiii}{[O\,{\footnotesize III}]}
\newcommand{\oiiit}{[O\,{\tiny III}]}

\newcommand{\sii}{[S\,{\footnotesize II}]}

\newcommand{\lir}{$L_{\rm WISE}$}

\newcommand{\feii}{Fe\,{\footnotesize II}}

\def\lax{{$\mathrel{\hbox{\rlap{\hbox{\lower4pt\hbox{$\sim$}}}\hbox{$<$}}}$}}
\def\gax{{$\mathrel{\hbox{\rlap{\hbox{\lower4pt\hbox{$\sim$}}}\hbox{$>$}}}$}}

\begin{document}


\title{CONNECTION BETWEEN MID-INFRARED EMISSION PROPERTIES AND NARROW LINE REGION OUTFLOWS IN TYPE~1 ACTIVE GALACTIC NUCLEI}
\author{ Kai~Zhang\altaffilmark{1, 2}, Ting-Gui Wang\altaffilmark{1}, Lin Yan\altaffilmark{3}, and Xiao-Bo~Dong\altaffilmark{1}}

\altaffiltext{1}{Key Laboratory for Research in Galaxies and
Cosmology, The University of Sciences and Technology of China,
Chinese Academy of Sciences, Hefei, Anhui 230026, China}
\altaffiltext{2}{Key Laboratory for Research in Galaxies and Cosmology, Shanghai
Astronomical Observatory, Chinese Academy of Sciences, 80 Nandan
Road, Shanghai 200030, China}

\altaffiltext{3}{Spitzer Science Center, California Institute of Technology,
1200 E. California Blvd., Pasadena CA 91125, USA}

\email{zhangkai@shao.ac.cn}
\shorttitle{NLR outflows and warm dust}
\shortauthors{Zhang et al.}
\begin{abstract}

The location of warm dust producing the Mid-infrared (MIR) emission in Type 1 Active Galactic Nuclei (AGNs) is
complex and not yet fully known. We explore this problem by studying how
the MIR covering factor ($CF_{\rm MIR}$ = $L_{\rm MIR}/\lbol$) correlates with
the fundamental parameters of AGN accretion process (such as $\lbol$, black hole mass \mbh, and
Eddington ratio \lratio)
and the properties of narrow emission lines (as represented by \oiii\,$\lambda5007$),
using large data sets derived from the Sloan Digital Sky Spectroscopic Survey (SDSS) and
the Wide Infrared Sky Survey (WISE).
Firstly we find that the luminosity of the \oiii\ wing component ($L_{\rm wing}$) correlates more tightly
with the continuum luminosity (\lfive) than the luminosity of the line core component ($L_{\rm core}$) does,
which is in line with our previous conclusion that the wing component, generally blueshifted,
originates from the polar outflows in the inner narrow-line region (NLR).
We then find that the MIR $CF$
shows the strongest correlation with $L_{\rm wing}/\lbol$
rather than with $L_{\rm core}/\lbol$ or the above fundamental AGN parameters, and the correlation
becomes stronger as the infrared wavelength increases.
We also confirm the anti-correlations of $CF_{\rm MIR}$ with \lbol\ and \mbh,
and the lack of dependence of $CF_{\rm MIR}$ on the Eddington ratio.
These results suggest that a large fraction of the warm dust producing MIR
emission in AGNs is likely embedded in polar outflows in the NLR instead of in the torus.
\end{abstract}


\section{Introduction}
The mid-infrared (MIR) emission of radio quiet active galactic
nuclei (AGNs) may come from the dusty torus and/or narrow
line region (NLR), as well as star formation regions in the host galaxies.
High-resolution imaging in the near-infrared (NIR) provides direct
evidence for the torus emission in a few nearby AGNs (Bock et al. 2000;
Radomski et al. 2003). At longer wavelengths, observations showed
extended MIR emission regions, coincident with the inner NLR, in addition
to the unresolved core, presumably the optically thick torus
(Tresch-Fienberg et al. 1987; Braatz et al. 1993; Cameron et al. 1993; Packham
et al. 2005; H\"{o}nig et al. 2010). Besides, MIR spectroscopic analysis revealed an
emission component, dominated by PAH emission, from star formation
regions, and its contribution increases with increasing
wavelength and decreasing AGN luminosity (e.g. Deo et al. 2009).

Recent high-resolution observations with the $MIDI$ instrument on $VLT$
demonstrated that in NGC1068, NGC 424, the Circinus galaxy
and probably also in NGC 3783,
most of the MIR emission lies along the polar direction,
orthogonal to the major-axis of the NIR emission region (Raban et al. 2009;
H\"{o}nig et al. 2010, 2012; Tristram et al. 2012). H\"{o}nig et al. (2012) proposed that the polar dust
is entrained in radiatively driven dusty winds, originated from the outerpart
of the accretion disk (e.g. K\"onigl \& Kartje 1994; Keating et
al. 2012), or evaporated from the dusty torus (Dorodnitsyn et al. 2008).
Thus the strength of MIR emission may be closely connected to the properties
of the gas outflow. The outflow in the NLR is manifested by the blue-shifted and relatively broad
component in the high-ionization narrow emission lines (e.g. Zhang et~al. 2011).
The NLR in nearby galaxies has been successfully modeled
by biconical outflows on scales of a few to tens of parsecs from the nuclei
in nearby Seyfert galaxies (Heckman et al. 1981; Das et al. 2005; Crenshaw et~al. 1999, 2010).
It is well known that the NLR contains dust.
It would be interesting to explore if
the MIR properties are linked to those of narrow emission lines.


The Wide-field Infrared Survey Explorer (WISE; Wright et al. 2010) surveyed
the whole sky at 3.4, 4.6, 12 and 22 $\mu$m at sensitivities about 100-1000
better than previous all sky surveys. Taking the advantage of the WISE and
the Sloan Digital Sky Survey (SDSS) data releases, it is possible to construct a
large sample of AGNs with both
measurements in MIR and optical to explore the connection between the MIR
properties and the AGN fundamental parameters or NLR properties. In what
follows, we adopt a $\Lambda$CDM cosmology with $H_{\rm 0}$ = 70 km\,
s$^{-1}$\,Mpc$^{-1}$, $\Omega_{\rm m}$ = 0.3, and $\Omega_{\rm \Lambda}$ = 0.7.


\section{The Sample and the Measurement}

\subsection{The Sample and Spectral Fit}
From the spectroscopic data set of the SDSS Data Release 4 (Adelman-McCarthy et al. 2006),
we select 4178 Seyfert 1 galaxies and quasars (namely type~1 AGNs) that have
reliable measurements of emission line fluxes and the continuum
suffers little host galaxy contamination. The sample is described
in detail in Dong et al. (2011).
As described in Dong et al. (2008), we fit simultaneously the continuum
and emission lines in the spectral regimes near H$\beta$+\oiii\ and H$\alpha$+\sii\
with a model consisting of several components: a power-law representing the
featureless AGN continuum, the \feii\ templates, and a number of gaussians
for emission lines. Each of the narrow emission lines is fitted with a single
Gaussian, except for the \oiii\,$\lambda \lambda$4959, 5007 doublet lines,
which are modeled with two Gaussians. One accounts
for the line core and the other for a possible blue wing.
The details of the fitting algorithm and spectral parameters can be found
in Dong et al. (2011) and Zhang et al. (2011).


\subsection{Mid-Infared Data}

We cross-correlate our SDSS Type 1 AGN catalog with the WISE All Sky Data Release
(Wright et al. 2010) catalog using a match radius of 3 arcseconds to construct
the mid-infrared detected type 1 AGN sample. Nearly all sources in our sample
are detected by WISE (4174 out of 4178). We refer to the WISE Vega magnitude
as $W1$, $W2$, $W3$ and $W4$ for the 3.4, 4.6, 12, and 22$\mu$m bands, respectively.
We reject the sources that have signal to noise ratio less than 3 (S$/$N$<3$) in any WISE band.
Radio loud AGNs are rejected
according to the catalog by Lu et al. (2010) to avoid contamination to \oiii\ flux and
continuum flux. The final sample has 3336 sources, and we call it 'extended sample' hereafter.
We further make a redshift cut of z$<$0.3 to minimize
potential cosmological evolution and uncertainty in the $k$-correction
that might complicate the correlation analysis.  This 'base sample' consists
of 608 sources. The magnitude in each band is converted to $\nu F_{\nu}$ according to the published
WISE zero points, and the $k$-correction is made by interpolating the Spectral Energy Distribution
(SED) at rest-frame wavelengths of 3.4, 4.6, 12, and 17$\mu$m. The $k$-corrected luminosities in
each band are marked
as $L1$, $L2$, $L3$ and $L4$. For every source we connect the adjacent bands on the
$\log \lambda F_{\lambda}$ versus $\log \lambda$ diagram to approximate the MIR SED,
and integrate the SED from 3.4 to 17$\mu$m to obtain the total MIR luminosity, denoted as
\lir.


\section{Distinction between the \oiii\ Core and Wing Component}
Zhang et al. (2011) found that the equivalent width (EW) of the core component
of \oiii\ correlates negatively with \lfive\ (Baldwin effect) while the wing component
displays no or a weak anti-Baldwin effect. The different behaviors suggest
that the two components come from distinct gas components regulated by
different physical processes. The gaseous kinematics of the \oiii\ core traces the motion of stars
in the gravitational potential well of bulges (e.g. Nelson et al. 2000;
Wang \& Lu 2001; Komossa et al. 2008), indicating that the \oiii\ core is
emitted by the photo-ionized ISM in the galaxy bulge. This is supported
by imaging spectroscopic observations of nearby Seyfert galaxies, which show that
the gaseous and stellar kinematics are coupled (e.g. Dumas et al. 2008). On
the other hand, the wing is much broader than the core  but narrower than
broad lines, indicating that its emitting region is much closer to the
nuclei than the core component. Furthermore, the wings are usually
blueshifted up to 1000 km~s$^{-1}$, suggesting they are from an outflow component. High
spatial resolution spectroscopic observations of nearby Seyfert galaxies revealed
individual high velocity knots up to a velocity of 3000 km~s$^{-1}$ on scales from
parsecs to up a hundred parsecs (e.g. Crenshaw et al 2010).
All these observations suggest that the wing
component is formed in the outflow at a few to a few hundred parsecs from
the nuclei.

To examine the difference between the core and the wing in detail, we plot the
luminosities of the \oiii\ core and wing against \lfive\ in Figure~1 using
the extended sample. Obviously, both the core and the wing luminosities
correlate with \lfive. The best fit of a log-linear model yields a slope of
0.86$\pm$0.01 for the core component, and 1.14$\pm$0.01 for the wing component. These
slopes are consistent with the analysis of Zhang et al (2011) on the Baldwin
effect. However, the plot reveals an interesting feature that the correlation
is much tighter for the wing than for the core. Qualitatively, the $RMS$ scatters
to the best fit of the log-linear model are 0.29~dex for the wing, and 0.42~dex for the core.
The difference is significant according to $F-$test with a chance probability
$P<10^{-26}$. The $L_{\rm wing}$ correlates more tightly with the
the broad \ha\ luminosity, which used for bolometric luminosity estimation
in next section, than the $L_{\rm core}$ does for the basic sample.
Note that the decomposition into the wing and the core, and other measurements
introduce an uncertainty of about 0.15~dex in the EWs of the wing and
the core.\footnote{For every source, we use best-fit 2-Gaussian model of the \oiii\,$\lambda5007$ line
plus the measurement errors of the SDSS spectrum to
generate a new mock spectrum, then fit the mock \oiii\ line with
two Gaussians. Statistically in the extended sample,
the standard deviations of the difference between the output and input fluxes
are 0.12 and 0.15~dex for the core and wing components, respectively. }

The tighter correlation between the wing and the continuum luminosity indicates
that outflows do not change dramatically from object to object, while the
larger scatter in the core component suggests that the variations of ISM states
like density, density profile, ionization state etc., are pronounced.
Physically, the ISM gas in the bulge is regulated by
several physical processes, such as gas accretion, star-formation, stellar
wind and AGN feedback. The competition between different processes may lead
to very diverse physical conditions of the gas among AGNs, which
introduces a scatter in the correlation. Interestingly, Zhang et al. (2011)
found that the EW of the \oiii\ core correlates strongest with the outflow
velocity of the whole \oiii\ profile (the outflow velocities being negative
due to the outflow, hence the correlation),
suggesting the interplay of
the outflow and ISM of the galaxy bulge. This can be interpreted most directly
in terms of AGN feedback: {\em high velocity outflows blow away the ISM in the
host bulge}. Also, continuum variations may introduce a
larger scatter in the core component than in the wing component because the
strength of the core component depends on the long-term variability
while the strength of the wing component follows more closely the
continuum variations (Zhang et al. 2013).
The effect of the variability depends on the light curves over time scales of
$10^3$ years.

\section{Correlation between $CF_{\rm MIR}$ and AGN properties}

We use the ratios of mid-infrared to bolometric luminosities to quantify
the MIR $CFs$, i.e. the relative contribution of MIR to bolometric
luminosity.
The bolometric luminosity is estimated from the broad line \ha\ luminosity using
a bolometric correction factor of 170, which is combining \lbol=12.17$\times$\lfive\ in Richards et al. (2006) and
\lfive=14$\times L_{\rm H\alpha}$ in Stern \& Laor (2012).\footnote{The bolometric correction may depend on
the accretion rate and luminosity (e.g. Vasudevan \& Fabian 2007; Lusso et al. 2012).} %
This is similar to the $130_{\div2.4}^{\times2.4}$ adopted in Stern \& Laor (2012).
We do not use \lfive\ directly as an estimate for the bolometric luminosity, because
it may be subject to host contamination and the optical-UV slope gets bluer in
higher luminosity AGN (See Section 4.1 in Zhang et al. (2013) for details).
The median bolometric luminosity of our sample is $10^{45}$ erg~s$^{-1}$.
In Figure~2 we plot the histograms of mid-infrared luminosity to bolometric
luminosity for the individual wavelengths (Panels a to d) and for the total MIR
luminosity (Panel e). All histograms show a log-normal distribution. For $CF_{\rm WISE}$,
the distribution is centered at log $CF_{\rm WISE}$ = -0.37 (43\%) and has a dispersion
of 0.243~dex. To first order, this means a covering factor of 43\%,
and is consistent with the type 2 fraction of
$\sim$50\% at this luminosity in literature (e.g. Hao et al. 2005; Lawrence \& Elvis 2010).
However, one needs to caution that the type 2 fraction may not be equal to the covering
factor of warm dust. Anisotropy in primary continuum and IR emission, presence of
large scale obscuring material and hot dust that emits at wavelength shorter than 3.4$\mu$m lead to
deviation of one from another (Calderone et al. 2012; Roseboom et al. 2013; Ma \& Wang 2013).

The $CF_{\rm MIR}$ spreads over a dynamic range (the range it spans) of more than
1.8~dex at the 17$\mu$m band and the dynamic range
increases with increasing wavelength.
This trend is clearly seen in Figure~2 in panel (a) to (d) respectively.
The $\sigma$'s of  $CF_{\rm MIR}$ increase from 0.200~dex at 3.4$\mu$m to
0.298~dex at 17$\mu$m.
One possible reason of increasing dispersion with wavelength is that star-formation contributes a
large fraction of emission at longer wavelengths. As discussed at the beginning
of Section~5, star-formation would
contribute less than 10\% at 17$\mu$m, thus introducing a dispersion less than
0.04~dex. This alone is not enough to account for the 0.22~dex difference between
dispersions of $CF1$ and $CF4$. The intrinsic variance after subtracting
the measurement errors in IR luminosity still
show a difference of 0.20~dex.
There are two possibilities for this difference: 1) The 17$\mu$m emission is from the dust
torus but the fraction of warm dust in the torus varies drastically from object to object
due to either different degree of clumpiness or different optical depth of the torus.
2) There is an additional warm dust component, and the contribution from the additional
component varies.
As discussed in Section~5.1, the torus explanation is inconsistent with our findings,
and we argue below that the latter is plausible.

\subsection{Correlations between $CF_{\rm MIR}$ and fundamental AGN properties}
First, we analyze how $CF_{\rm MIR}$ correlate with redshift and the AGN fundamental
parameters, such as \lbol, \mbh\ and \lratio\ using
the basic sample. The results are summarized in column 1, 2, 3, 4 of Table~1.
We calculate the black hole masses based on the FWHM of broad \hb\ and \lfive\ using
the formalism presented in Wang et al. (2009, their Eqn.~11).
The Eddington ratio (\lratio) is the ratios between the bolometric and Eddington luminosities:
$\lratio=\frac{\lbol}{\ledd}$, $\ledd  = 1.26 \times 10^{38}$ (\mbh/\msun) erg~s$^{-1}$.
We use Spearman correlation coefficient ($\rho$) as a
non-parametric measure of the statistical dependence between two quantities.
\\

We find that the $CF_{\rm MIR}$ correlate negatively with \lbol\ and
\mbh, but very weakly with \lratio.
We get $CF1 \propto \lbol^{-0.21}$, $CF2 \propto \lbol^{-0.21}$, $CF3 \propto \lbol^{-0.31}$,
$CF4 \propto \lbol^{-0.35}$, and $CF_{\rm WISE}/ \propto \lbol^{-0.26}$ respectively as shown in Figure~3.
The typical error of the slopes is 0.03.
These results are fully consistent with previous works (Maiolino et al. 2007; Treister et al. 2008;
Calderone et al. 2012; Ma \& Wang 2013; Roseboom et al. 2013).
We plot the best regression fits in Calderone et al. (2012) and
Ma \& Wang (2013) on Panel (e) to illustrate the consistency.
Calderone et al. (2012) integrated
the MIR luminosity at 2.1-13.7$\mu$m, and Ma \& Wang (2013) used 3-10$\mu$m luminosity.
To make a fair comparison, we add 0.184~dex to the $CF_{\rm MIR}$ in Ma \& Wang (2013) to account
for the shorter wavelength range. The 0.184~dex is the mean ratio of 3.4-17$\mu$m
luminosity to 3.4-11.6$\mu$m luminosity in our sample. We do not make any correction
to the regression fit in Calderone et al. (2012) because the gain in 2-3.4$\mu$m approximately equals to
the loss in 13.7-17$\mu$m according to the mean SED of all quasars in Richards et al. (2006).
We can see that the fitting lines from literatures and our best regression fit
are consistent with each other at this luminosity range.
This further supports the bolometric correction we adopt is reasonable.

The slopes of $CF_{\rm MIR}$ versus \mbh\ relations are
-0.22, -0.23, -0.35, -0.42, -0.30 for L1, L2, L3, L4, and \lir\ respectively, consistent with previous works
(e.g. Mor et al. 2011; Ma \& Wang 2013).
The lack of dependence of $CF_{\rm MIR}$ on \lratio\ is reported by many authors (Cao et al. 2005;
Mor et al. 2011; Ma \& Wang 2013).
Considering the limited range of \lratio (1.5~dex) for our
sample, we refrain from over-interpreting this result.

\subsection{Correlations between $CF_{\rm MIR}$ and narrow emission line properties}
Next, we explore whether the MIR emission is linked to the properties of optical narrow
emission lines as represented by \oiii\,$\lambda5007$.
We present the results of correlation analysis between $CF_{\rm MIR}$ and
the \oiii\ total, core, wing luminosity to \lbol\ ratios using the basic sample.
The results are summarized in column 5, 6, 7 of Table~1.
The $CF_{\rm MIR}$ in all 4 bands correlate significantly with
$L_{\rm wing}$/\lbol.
Among them, the strongest one is between $CF4$ and $L_{\rm wing}$/\lbol\ with
$\rho=0.52$. The correlation weakens as wavelength decreases: with
$\rho=$0.47, 0.36, 0.32 between $L_{\rm wing}$/\lbol\ and $CF3$, $CF2$ and
$CF1$, respectively.
This is in parallel with the increasing dynamic range of the $CFs$ with wavelengths.
Note that the bolometric correction has $RMS$
scatters of order 0.38~dex. So it is likely that the weakening of the
correlations is a combination of decreasing dynamic range and relatively
large uncertainty in the bolometric correction. The physical reason of increasing
dynamic range with increasing wavelength is discussed in Section 5.3. We plot $CF4$ versus
$L_{\rm wing}$/\lbol\ in Figure~4. A log linear fit yields a slope of 1.19$\pm$0.04
with an $RMS$ scatter of 0.308~dex. This scatter is significantly larger than
the combinations of measurement errors: 0.11~dex in L4/$L_{\rm H\alpha}$ and 0.15~dex
in $L_{\rm wing}$/$L_{\rm H\alpha}$, indicating that there is a considerable intrinsic scatter
in this relation. No correlation is found between outflow velocity and any IR
properties in our sample.

\section{Discussion}

To facilitate discussions on the implication of the correlation between
$L_{\rm wing}$/\lbol\ and L4/\lbol, we
note that the star-formation contamination is likely small in the MIR band for
these luminous type 1 AGNs in our sample. For most of our sources, the
infrared luminosity at 17$\mu$m is in the range of  $10^{43}$ to
$10^{46}$ erg~s$^{-1}$ ($10^{9.5}$ to $10^{12.5}L_{\bigodot}$).
According to Fig~10 of Deo et al. (2009), starburst contribution is
likely less than 10\% at $17\mu$m and a few percent in the L1 band.
This is more likely to be true because we have removed any source with substantial
stellar contamination in optical.
In the following sections, we will discuss the location of the
warm dust and its relation with torus in light of the above results.

\subsection{Classical Steady Torus?}
Judging from studies of AGN at subarcsecond resolution in the MIR, the majority
of MIR emission originates from the central few pc region (Jaffe et al. 2004;
Tristram et al. 2007). In NGC 1068, about 75\% of the MIR emission is from
the central unresolved core (Bock et al. 2000). This was believed to be the
torus, and a variety of torus models were constructed to match the NIR--MIR
SED (e.g. Fritz et al. 2006; H\"{o}nig et al. 2006; Schartmann et al. 2008;
Nenkova et al. 2008a, b).
If the dust dominates the MIR emission,
as the torus acts to confine the ionization cone,
there should be an anti-correlation between the torus subtending angle and
the covering factor of NLR; we thus should obtain a negative correlation between
$CF4$ and $L_{\rm core}$/\lbol.
This is however in contradiction to the observation.

\subsection{Dusty NLR clouds?}

NLR is theoretically expected to contain dust (Netzer \& Laor 1993; Dopita et al. 2002).
Observationally, extended polar MIR emission is found around the compact cores of
several AGNs (See Section~1).
Groves et al. (2006) modeled the narrow line spectrum of NGC 4151
with dusty photo-ionization models, and found that 10--50\% of its 25$\mu$m
emission is from the NLR dust. The imaging observations also indicate
that $>$27\% of MIR emission emerges from a region coincident with the NLR
in NGC 4151 (Radomski et al. 2003). The warm dust extends from the unresolved
core to about 100pc. Mor et al. (2009) fitted the MIR
spectra of PG QSOs with 3 components: a black body, the clumpy torus
emission (Nenkova et al. 2008a,b) and the NLR dust. They derived that above 18$\mu$m,
the NLR dust contributes as much as 40\% of the emission.
Schweitzer et al. (2008) demonstrated
the the NLR dust could produce the silicate emission features observed in AGNs.
As the cold gas and the dust are coupled in the NLR, this would naturally explain the
correlation between $CF4$ and $L_{\rm wing}$/\lbol\, and the correlation
weakens at shorter infrared wavebands, qualitatively.

However, the NLR dust models do not consider the dust in the outflow
component of NLR, so they can not explain why $CF4$ is much more
strongly correlated with $L_{\rm wing}$/\lbol\ than with $L_{\rm core}$/\lbol.
Furthermore, most of the infrared emission from 5 to 18$\mu$m is attributed to
the torus emission, which is not consistent with more recent higher
resolution imaging observations in several nearby Seyfert galaxies (see \S 1).
In Mor et al. (2009),
the distance of NLR dust is estimated to be $R_{NLR}=295\times L_{46}^{0.5}$ pc,
where $L_{46}$ is the bolometric luminosity in unit of $10^{46}$ erg~s$^{-1}$.
In high luminosity sources, we lack observational evidence that the outflow
may reach that far from the nuclei.
Although detailed model is needed to predict the infrared spectrum, it is clear
that any optical thin dust in the outflow region will be hotter than that
in the more distant classical NLR, and will contribute significantly to emission
below 15 $\mu$m. Alternatively, the dust clouds in outflows
are optically thick to infrared light, radiative transfer effect would result in
much colder infrared emission. We therefore do not think that dusty NLR clouds
on scales of $\gtrsim$100 pc are responsible for our observed correlation of
$L_{\rm wing}/\lbol$ with $CF_{\rm MIR}$.

\subsection{Dusty Outflows?}
It was pointed out by Krolik and Begelman (1988) that a steady doughnut torus
suffers the problem that there is not enough vertical motions capable of
sustaining the H/R$\sim$1 geometry. One solution to this problem is assuming that
the dust is entrained in the magnetically driven accretion disk wind
(Emmering, Blandford \& Shlosman 1992; Elitzur \& Shlosman 2006),
and the outer part of the outflow acts as a clumpy torus.
Alternatively, the IR radiation pressure on dust may also offer enough pressure
to support a thick torus, even launches an outflow (Krolik 2007;
Dorodnitsyn et al. 2011, 2012). In this type of dynamical models, the outflow
and the torus are naturally linked. A connection between the outflow and the dust emission
is expected.
The detection of dominant MIR emission in polar direction led H\"{o}nig et al.
(2012) to propose that the polar dust is entrained in a radiatively-driven dusty
winds. This dust component is an inward extent
of NLR dust observed in many nearby AGNs but lies only a few pc away from the nuclei.
This is where the outflow originates and accelerated. In this paradigm, the outflow gas
and the warm dust share similar covering factor, and would lead to our finding
of a positive correlation between $CF4$ and $L_{\rm wing}$/\lbol\ naturally.
The contribution of the polar warm dust to the mid-infrared emission
increases with wavelength, so the correlation between
$CF_{\rm MIR}$ and $L_{\rm wing}$/\lbol\ strengthens with wavelength.
By comparison, the static classical NLR traced by the \oiii\ core component lies further
from the nuclei. The dust embedded in the static NLR emit dominantly at wavelength
longer than 20$\mu$m, thus only connects to the warm dust loosely.
If the correlation between the warm dust and the outflow strength is really
driven by the dust in the polar winds, an implication is that
the pc scale MIR elongation in NGC 1068, the Circinus galaxy and NGC 424 may not be special
cases, but a common phenomenon for Seyfert galaxies. Most of the warm
dust may reside in the outflow instead of in the torus.
Besides, the outflow may have a significant
effect on the dust re-distribution by transporting the dust from the central region
to larger scales.

\acknowledgements
We thank the referee for helpful suggestions that improve the paper significantly.
This work is supported by Chinese NSF grants
NSFC 11233002, 11073019, and a Chinese Universities Scientific Fund (CUSF) for USTC.
This publication makes use of data products
from the Wide-field Infrared Survey Explorer, which is a
joint project of the University of California, Los Angeles, and
the Jet Propulsion Laboratory/California Institute of Technology,
funded by the National Aeronautics and Space Administration.
This study makes use of data from the SDSS (see:
http://www.sdss.org/collaboration/credits.html).

\begin{table*}[h]
\topmargin 0.0cm
\evensidemargin = 0mm
\oddsidemargin = 0mm
\scriptsize 
\caption{Correlation Tables between the IR covering factors and the optical parameters \tablenotemark{a}}
\label{corrtab}
\medskip
\vfill
\begin{tabular}{l|c c c c c c c c }
\hline \hline
            & z          & \lbol     &   \mbh       & \lratio    & $L_{\oiiit}$/\lbol &  $L_{\rm core}$/\lbol & $L_{\rm wing}$/\lbol          \\
\hline
$CF_{\rm WISE}$ & 0.06(1e-01) & -0.40(3e-25) & -0.38(2e-22) & -0.10(7e-03) & 0.28(1e-12) &  0.09(2e-02)& 0.45(1e-32) \\
$CF1$  &  0.08(2e-02) & -0.37(5e-22) & -0.32(8e-17) & -0.12(1e-03) & 0.10(1e-02) & -0.07(6e-02)& 0.32(3e-16)  \\
$CF2$  &  0.09(1e-02) & -0.35(4e-20) & -0.33(2e-17) & -0.09(1e-02) & 0.14(5e-04) & -0.04(2e-01)& 0.36(8e-20) \\
$CF3$  &  0.05(1e-01) & -0.40(4e-25) & -0.38(5e-23) & -0.10(1e-02) & 0.31(2e-15) &  0.12(1e-03)& 0.47(4e-35)  \\
$CF4$  &  0.02(5e-01) & -0.42(2e-27) & -0.40(3e-25) & -0.10(6e-03) & 0.43(9e-29) &  0.25(9e-11)& {\bf 0.52}(1e-43) \\
\hline
\end{tabular}
\medskip
\vfill
{\normalsize $^a$~For each entry, we list the Spearman rank correlation
      coefficient ($\rho$) and the probability of the null hypothesis that
      the correlation is not present (\pnull) in parenthesis,
      for the 608 objects in the basic sample.$\rho \geq$0.50 are marked in {\bf Boldface}. }\\
\end{table*}


\begin{figure*}
\begin{center}
\label{fig-1}
\includegraphics[width=16cm]{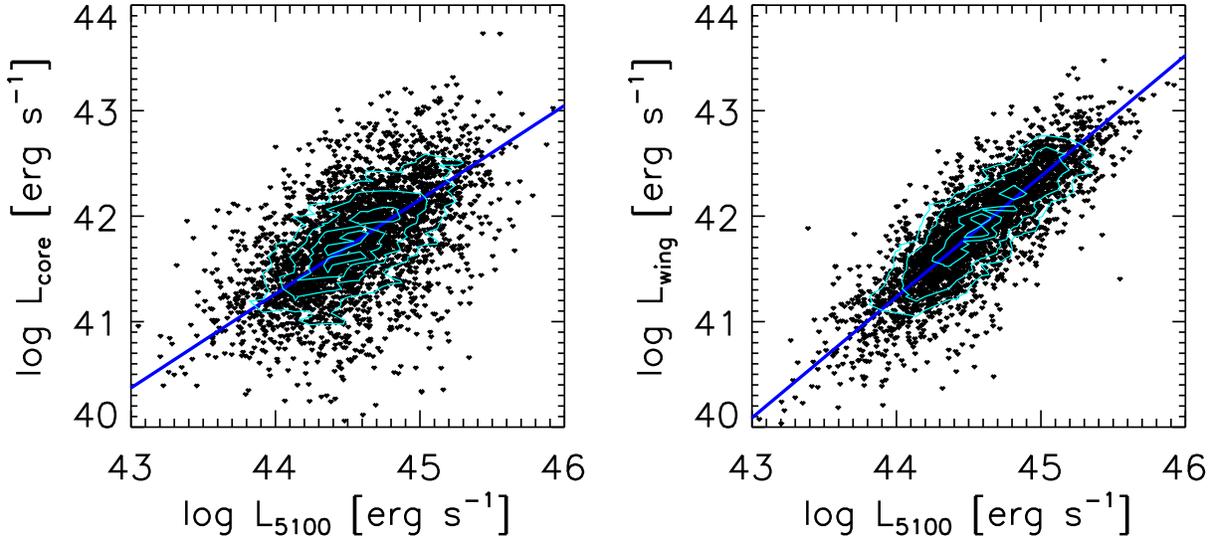}
\caption{Left: The \oiii\ core
luminosity versus \lfive\ using the extended sample (3336 sources).
The solid line shows the best fitting line, and the cyan contours
show the density of the distribution. The levels of the contours are
15, 30, 45, 60 and 75 respectively.
Right: The \oiii\ wing
luminosity against \lfive\ using the extended sample. The levels of the
contours are 15, 30, 60 and 80. }
\end{center}
\end{figure*}

\begin{figure}
\begin{center}
\label{fig-1}
\includegraphics[width=16cm]{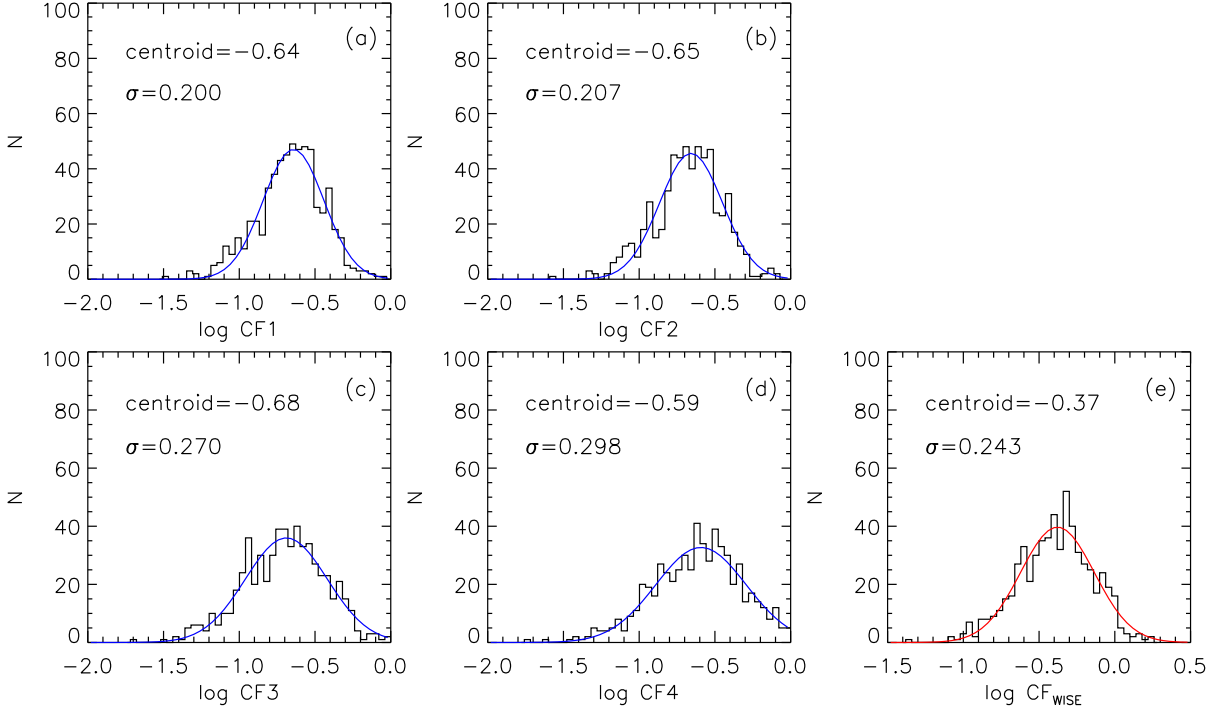}
\caption{Distributions of MIR covering factors: $CF_{\rm MIR}$ = $L_{\rm MIR}/\lbol$ using the basic sample.
The solid line is the gaussian fit to the
distribution, the centroid and $\sigma$ of the gaussian are shown in the left-up corner. Panel (a) to (d)
are $CF_{\rm MIR}$ distributions at each band, and Panel (e) is the distribution of $CF_{\rm WISE}=L_{\rm WISE}$/\lbol. }
\end{center}
\end{figure}

\begin{figure}
\begin{center}
\label{fig-1}
\includegraphics[width=16cm]{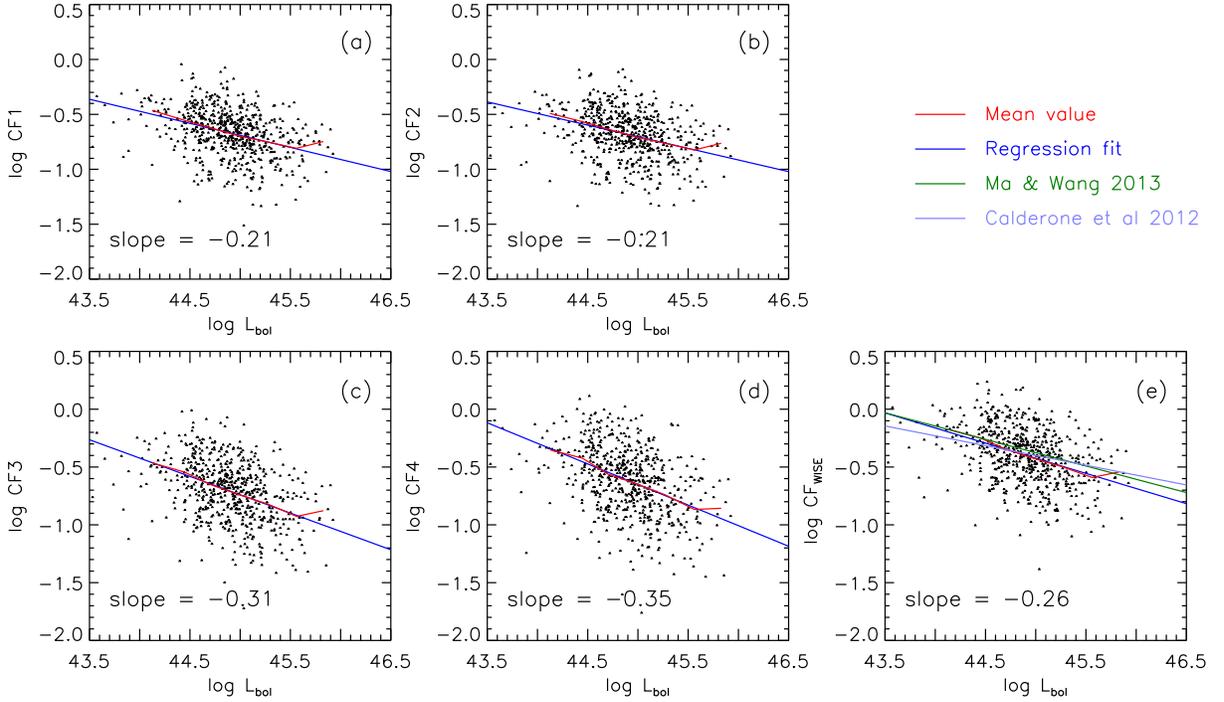}
\caption{$CF_{\rm MIR}$ against \lbol\ of the basic sample.
The blue line is the best regression fit, and the red line
is the mean value in each \lbol\ bin with binsize of 0.3~dex starting from $\lbol=10^{43.8} erg$ $s^{-1}$. The
slope of the regression fit is shown in the left-down corner. Panel (a) to (d)
are $CF_{\rm MIR}$ versus \lbol\ at each band, and Panel (e) is the $CF_{\rm WISE}$ versus
the bolometric luminosity. The green and purple lines are the results from Ma \& Wang (2013) and Calderone et al. (2012).}
\end{center}
\end{figure}

\begin{figure}
\begin{center}
\label{fig-1}
\includegraphics[width=8cm]{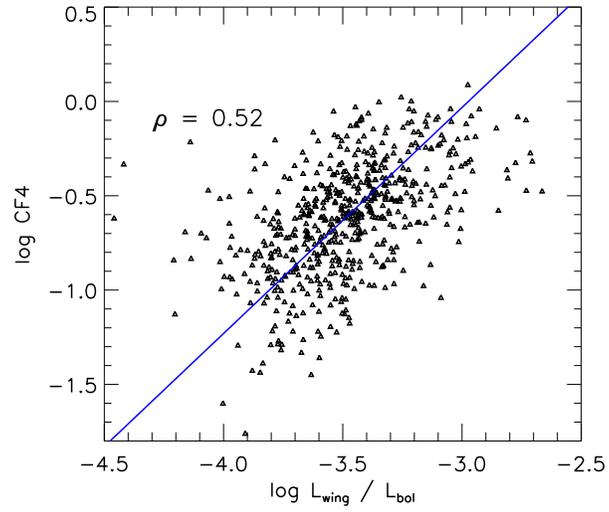}
\caption{$CF4$ versus $L_{\rm wing}/\lbol$ using the basic sample (608 sources). The correlation
coefficient is shown in the left-up corner. The solid line is the best regression fit.  }
\end{center}
\end{figure}

\clearpage

\end{document}